\documentclass{xeus}
\usepackage{graphics}
\usepackage{psfig}

\begin{document}
\onecolumn
\title{The X-ray evolving universe:
(ionized) absorption and dust,  from nearby Seyfert galaxies
to high-redshift quasars}
\author{Stefanie Komossa, G\"unther Hasinger }
\institute{Max--Planck--Institut f\"ur extraterrestrische Physik,
 Giessenbachstra{\ss}e, 85748 Garching, Germany
}
\titlerunning{(Ionized) absorption and dust from low to high redshift}
\maketitle

\begin{abstract}

Cold and warm absorbers
have beeen detected in all types of active galaxies (AGN) 
from low to high redshift.  This gas, located in the black hole
region of AGN, is thought to play an important
role in AGN unification scenarios, in explaining the
X-ray background, in black hole growth
and AGN evolution.

High-resolution spectroscopy with {\sl Chandra} and {\sl XMM-Newton}
has recently revealed the signatures of warm absorbers in the form
of many narrow absorption lines from highly ionized material.
The richness in spectral features will provide a
wealth of information on the physical processes in the
central region of the few X-ray brightest, most nearby Seyfert galaxies.
The long-term goal is to obtain this information for a much
larger number of objects, particularly at higher redshift. 
This will be possible with the future X-ray observatory {\sl XEUS}.    

We provide a review of the observations of dusty and dust-free warm and cold
absorbers at low and high redshift, including most recent results
and exciting questions still open.  Emphasis is on the
science issues that we will be able to address with {\sl XEUS}
for the first time, particularly at high redshift, including:
(i) determination of metal abundances
of X-ray (cold) absorbers by
detection of metal absorption edges,
(ii) analysis of the composition of dust mixed with cold and ionized
gas (K-edges of metals in cold dust and cold gas will be resolvable from each other for the first time),
(iii) measurement of the velocity field of the gas,
(iv) utilization of these results to investigate the {\em evolution} of gas and dust
in AGN from high to low redshift: the evolution of abundances, dust content,
ionization state,
amount and velocity of gas, and its role in feeding the black hole.

We emphasize the importance of iron absorption 
measurements with {\sl XEUS}
at high redshift for two key issues of cosmology:
the early star formation history of the universe, and 
the measurement of cosmological parameters.  
As an example, we discuss recent {\sl XMM-Newton} observations of the
high-redshift BAL quasar APM 08279+5255. 

\end{abstract}

\section{\large Introduction}

Neutral (`cold') or ionized (`warm') gaseous material is ubiquitous in
the AGN/SMBH environment, and therefore of utmost importance in
understanding the AGN phenomenon, 
the evolution of active galaxies,
their link with starburst galaxies and ULIRGs, and the X-ray background.
X-ray absorption and emission features provide
valuable diagnostics of the physical conditions
in the X-ray gas and, in particular, allow to measure elemental
abundances at high redshift, with profound consequences for
our understanding of the star formation history in the early universe. 

Here, we provide a short overview of previous X-ray observations
of absorption in AGN, and discuss how exciting questions still open can be
addressed with the {\sl XEUS} observatory. 
We apologize in advance for incompleteness in citations due to space limitations.  

\section{\large (Warm) absorbers in nearby Seyfert galaxies}

\subsection{Dust-free and dusty ionized absorbers}

With {\sl ROSAT}, the signatures of warm absorbers, absorption edges
of highly ionized oxygen ions at
$E_{\rm OVII}=0.74$ keV and $E_{\rm OVIII}=0.87$ keV,
were first detected in the X-ray spectrum of MCG$-$6-30-15
(Nandra \& Pounds 1992), following earlier {\sl Einstein} evidence for
highly ionized absorbing material in AGN
(Halpern 1984). Detailed studies of many
other AGN followed, and the signatures of warm absorbers have now
been seen in about 50\% of the well-studied
Seyfert galaxies (e.g., George et al. 1998; see Komossa 1999 for a review).
First constraints placed the bulk of the ionized material outside
the BLR (e.g., Mathur et al. 1994, Komossa \& Fink 1997b). 
Depending on its covering factor and location, the warm absorber
may be one of the most massive components of the active nucleus.

Some (but not all) ionized absorbers were suggested to contain dust,
based on otherwise contradictory optical--X-ray observations
(e.g., Brandt et al. 1996, Komossa \& Fink 1997b, Komossa \& Bade 1998).
The first possible direct detection of Fe-L dust features in the X-ray spectrum
of MCG$-$6-30-15 was recently reported by
Lee et al. (2001).
See Komossa (1999) for a pre-{\sl Chandra/XMM} review on warm absorbers.

{\sl Chandra} and {\sl XMM-Newton} detected a wealth of absorption 
features originating from ionized gas in nearby Seyfert galaxies
(e.g., Kaastra et al. 2000,
2002; Kaspi et al. 2000, 2001;
Sako et al. 2001; Collinge et al. 2001; Branduardi-Raymont et al. 2001;
Lee et al. 2001; Komossa et al. 2001; Netzer et al. 2002;
Yaqoob et al. 2002). These observations
confirmed the presence of ionized absorbers, but also showed that
spectra are more complex than previously modeled: 
The ionized absorbers are often multi-component, with a range
of ionization parameters and outflow velocities (e.g., Kaastra et al. 2002,
Kaspi et al. 2001),
 and absorption-dips
previously mainly attributed to oxygen edges are in some cases
similarly well or better explained by Fe-M absorption-line complexes
(e.g., Sako et al. 2001, Behar et al. 2002).
In the case of MGC\,-6-30-15 a discussion has started on how 
much of the ``saw-tooth'' spectral structure at soft X-ray energies originates
from a warm absorber (which is independently detected
by narrow absorption lines; Branduardi-Raymont et al. 2001,
Lee et al. 2001), or whether it is dominated by 
relativistically broadened accretion-disk lines 
(Branduardi-Raymont et al. 2001, Fabian
2001).  
%
\begin{figure}[b]
\hspace*{0.3cm}
\psfig{file=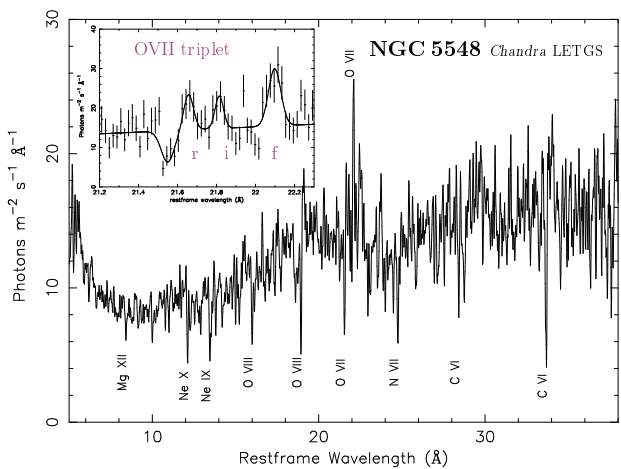,width=8.5cm,clip=}
\hfill
\begin{minipage}[]{0.42\hsize}\vspace*{-4.0cm}    
\hfill
\caption{High-resolution X-ray spectrum of NGC\,5548 (Kaastra et al. 2000)
obtained with the LETGS aboard {\sl Chandra}. 
The inset shows a zoom of the OVII triplet to which
a resonance line, two intercombination lines (unresolved), 
and a dipole-forbidden line
contribute.}
\end{minipage} 
\label{wa}
 \end{figure}
%

The presence of ionized absorbers may also be responsible for a number
of X-ray phenomena in X-ray-weaker objects, where previous X-ray
spectroscopy did no longer allow to resolve individual
spectral features from the absorber, but the collective
effect of the absorbing gas on the X-ray spectrum is still visible. 
We give two examples: 
for instance, (i) with {\sl ROSAT} it was generally
not possible to distinguish between black bodies and
ionized absorbers to account for the very {\em steep}
observed soft X-ray spectra of Narrow-line Seyfert-1 galaxies 
(Komossa \& Fink 1997a).  
(ii) The idea that {\em variations} of the
ionized absorber in response to intrinsic luminosity variations
can mimic continuum-shape variations in case of insufficient spectral
resolution is an old one, and was discussed in the 
early days of warm-absorber studies
(e.g., Kunieda et al. 1992).
More recently, it was applied to some cases of unusual variability
among AGN: the narrow-line Seyfert\,1 galaxy RXJ0134-4258
(Komossa \& Meerschweinchen 2000) and the Seyfert galaxy NGC\,3516
(Netzer et al. 2002).

\subsection{Cold absorption}
Cold absorption plays a fundamental role in Seyfert galaxies of type 2.
In this field, the {\sl BeppoSAX} mission recently led to great progress: 
large X-ray absorption columns were measured in many 
Seyfert\,2 and intermediate-type Seyfert galaxies, including a number of Compton-thick
candidate sources (e.g., Maiolino et al. 1998, Guainazzi et al. 1999, Bassani et al. 1999).  
According to Bassani et al., the mean absorption column in Seyfert\,2 galaxies
is about $N_{\rm H}  \simeq 10^{23.5}$ cm$^{-2}$, while about 20-30\% of the sources
of their sample exceed $N_{\rm H} > 10^{24}$ cm$^{-2}$.  
The X-ray absorption columns appear to be variable on the timescale
of months to years (e.g., Risaliti et al. 2002). Interestingly, some 
Seyfert galaxies seem to change their X-ray spectra from reflection-dominated
to transmission-dominated within several years
(Guainazzi et al. 2002, and ref. therein).  
For the relevance of cold absorption in the context of models for the
X-ray background, see Hasinger (these proceedings).

\subsection{Open questions which we will be able to address with {\sl XEUS}} 

{\sl Chandra} and {\sl XMM-Newton} provided high-quality spectra 
for the nearest Seyfert galaxies.  
The long-term goal is to obtain this information for a much
larger number of objects, particularly at higher redshift.
High spectral resolution and sensitivity will allow to detect
even weak lines, measure line-profiles, resolve multiple components,
perform line-reverberation mapping in X-rays, and obtain the velocity
fields of the absorber(s).  

Questions of particular interest are: 
how many components are warm absorbers composed of ?,
what are their densities, locations, covering factors, and metal abundances ?,
 is the ionized material in or out of photoionization equilibrium ?,
is the velocity field similar to that measured in the UV ? 
Finally, are UV- and X-ray absorber identical (e.g., Mathur et al. 1997,
Brandt et al. 2002)?, what
is the origin and evolution of ionized absorbers ? 
 
Concerning {\em dusty} warm absorbers, dust-created 
metal K-shell absorption edges will be spectrally resolvable from 
gas-created K-shell edges for the first time. Measuring dust
absorption in X-rays will be a powerful new tool to determine
the dust composition in other galaxies (e.g., Komossa \& Bade 1998,
Komossa 1999).  

Finally, all kinds of peculiar or extreme properties of AGN, partly suggested
to be linked to ionized absorption (e.g., Komossa \& Meerscheinchen 2000),
can be studied with {\sl XEUS} observations of high resolution and
sensitivity. 

\begin{figure}[b]
\hspace*{0.3cm}
\psfig{file=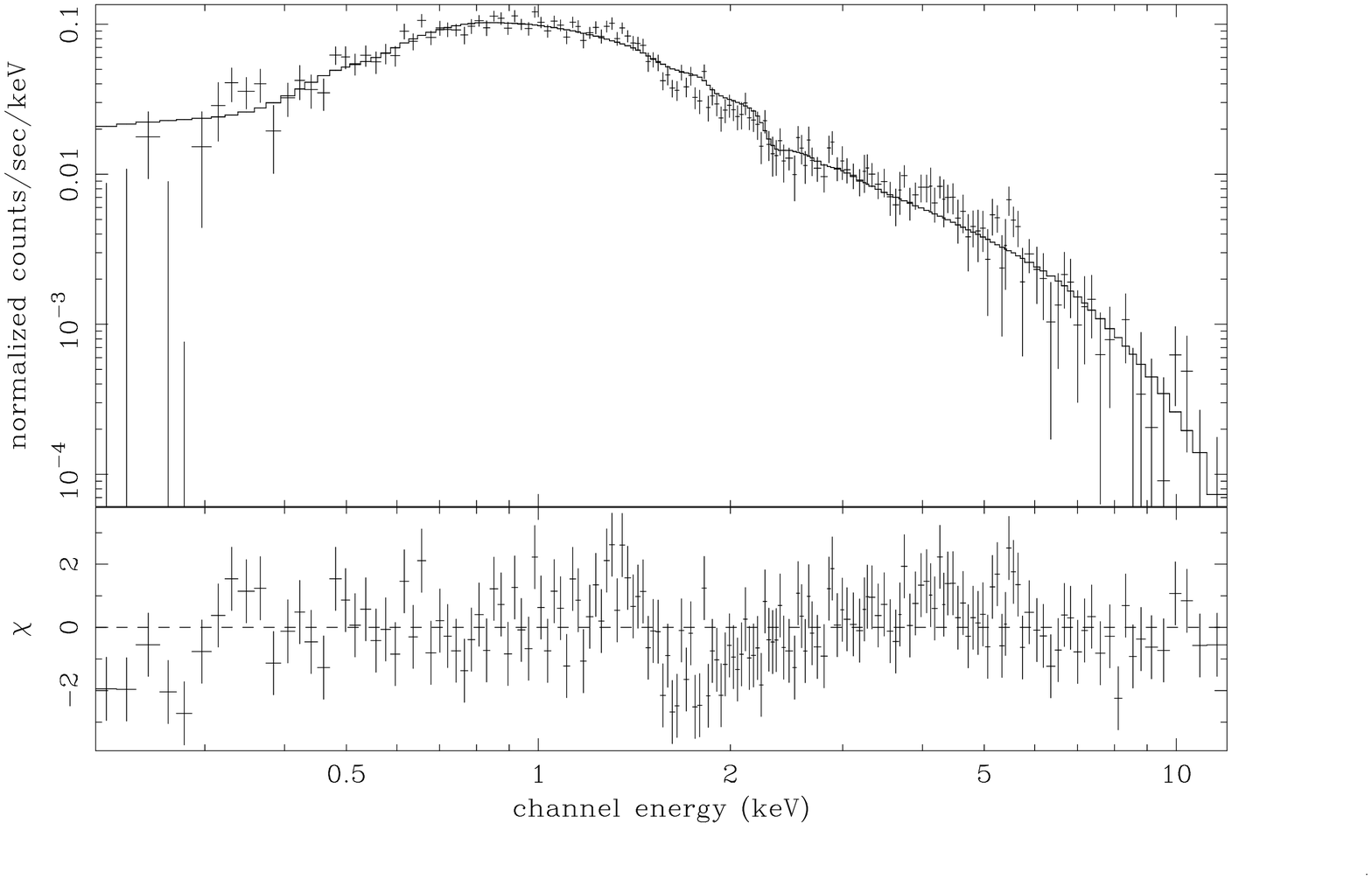,width=8.6cm,clip=}
\hfill
\begin{minipage}[]{0.48\hsize}\vspace*{-6.35cm}    
\hfill
\psfig{file=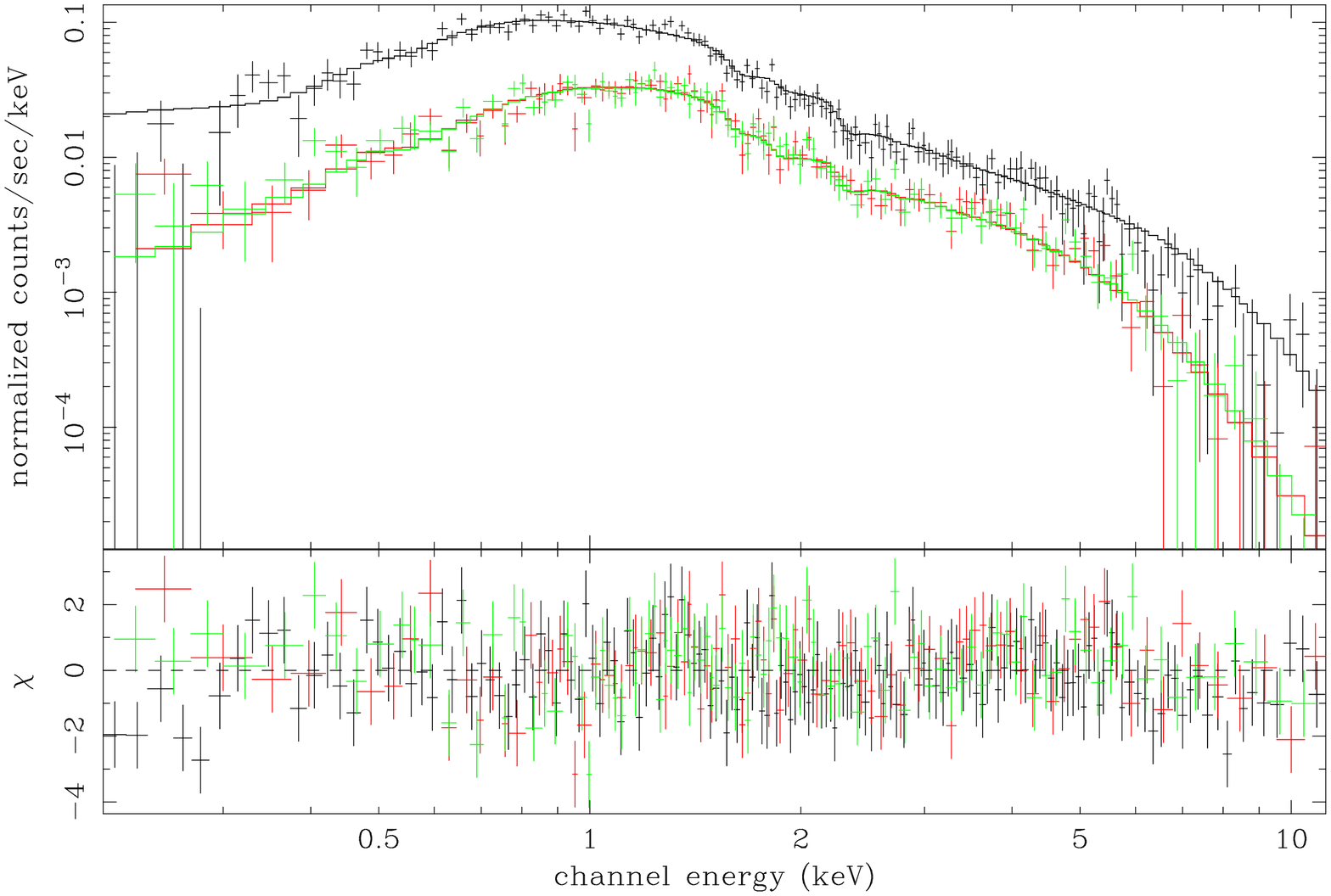,width=8.6cm,clip=}
\end{minipage}
\vspace*{-0.5cm}
\caption{ {\sl XMM-Newton} spectrum of the BAL quasar APM 08279+5255
at redshift $z$=3.91 
(Hasinger et al. 2002). {\em Left:} {\sl XMM} EPIC-pn spectrum,
fit with a single powerlaw. An absorption feature
is visible at an energy corresponding 
to ionized, redshifted iron.
{\em Right:} Combined EPIC and MOS spectra, fit with
a powerlaw plus an absorption edge of highly ionized iron.
}
\label{wa}
 \end{figure}

\section{\large Broad absorption line (BAL) quasars} 

\subsection{Previous observations}

BAL quasars are characterized by broad UV absorption lines.
Is has been suggested that these lines arise in
a flow of gas which rises vertically from
a narrow range of radii from the accretion disk. The flow then bends
and forms a conical wind moving radially outwards (Elvis 2000). 
Variants of radiatively-driven disk-winds were explored
(e.g., Murray et al. 1995, Proga et al. 2000, Proga 2001, Everett et al. 2002).  
In some of these models, an X-ray absorber shields the wind downstream
from soft X-rays, allowing resonant-line driving to remain effective and
accelerate the outflowing BAL wind up to $\sim$0.1c.  

Pre-{\sl Chandra/XMM} detections of BAL quasars in X-rays were rare. 
Generally,
BAL quasars are X-ray weak, which is usually interpreted in terms
of strong excess absorption
(e.g., Green et al. 1995, Gallagher et al. 1999, Brinkmann et al. 1999,
Brandt et al. 2000, Wang et al. 2000). 
Although {\sl Chandra} provided valuable new
constraints on the amount of absorption towards
selected BALs (e.g., Sabra \& Hamann 2001,
Oshima et al. 2001, Gallagher et al. 2002) almost all data still suffer from low S/N
(typically 50 to few hundred X-ray photons detected).
There are indications that the BAL material
is ionized instead of neutral.
This is definitely the case for the quasar APM 08279+5255 
which has the 
best-measured X-ray spectrum of any BAL quasar we are aware of. 
A recent {\sl XMM-Newton} observation led to the detection
of a strong absorption feature of ionized iron, interpreted as K-edge,
arising from a warm
absorber of high column density (Hasinger et al. 2002; for {\sl Chandra} results
on this quasar, just posted at astro-ph, see Chartas et al. 2002){\footnote{Chartas
et al. presented a {\sl Chandra} spectrum of APM 08279+5255 and modeled 
the absorption structure
by two iron absorption lines  
which then have huge outflow velocities of 0.2 and 0.4c. Both observations
are consistent with each other, if variability
is assumed.}}. 
No doubt, APM 08279+5255 is a top target for {\sl XEUS}
(see also Section 6).

\subsection{Open questions which will be addressed with {\sl XEUS}} 

One basic question related to the X-ray BAL flow is:
are we directly seeing the outflowing gas in X-rays, or
is the X-ray absorber at rest, shielding the UV absorbing
gas to ensure the line-driving remains effective even for
high ionization parameters ? Are UV and X-ray absorber
identical ? Is radiation pressure from UV lines indeed
the main driving mechanism of the outflow ?
What are the X-ray column densities and the 
corresponding  mass outflow rates ?  
What is the geometry of the BAL flow ? 

Of fundamental importance will be the simultaneous 
measurement of {\em column density} and {\em outflow velocity}
of the gas. (With present X-ray missions, and as long as the X-ray spectrum
is dominated by iron {\em absorption edges} it is very difficult to
distinguish between dominant ionization stage and outflow velocity of the gas.)
Such measurements 
will allow to solve one majour uncertainty in BAL models,
mentioned above: is the (high-column-density) X-ray absorber 
outflowing with high speed, or at rest ?
In fact, the high X-ray column density
measured in some BALs, most reliably for APM 08279+5255, 
{\em in combination with
high outflow velocities}  would
pose a problem for radiation-driven outflows (see, e.g., discussion by Hamann 1998) 
and may give indications that other mechanisms are
at work to drive the BAL flows.  
The new X-ray measurements with {\sl XEUS} 
will have profound implications for our understanding
of massive and energetic outflows in AGN, their launch and 
acceleration mechanism.

Abundance measurements will tell the history and origin of the BAL gas, 
and its role in metal-enriching the
environment.    
According to a recent model by Elvis et al. (2002),
BAL environments are expected to be dusty. In the X-rays regime,
absorption features from dust will provide a clean way
to measure the dust composition.

\section{\large Absorption in high-redshift quasars}

\subsection{Previous observations}

Evidence for excess X-ray absorption was found in  
high-redshift, mostly radio-loud, quasars
(e.g., Elvis et al. 1994, Schartel et al. 1997,
Yuan 1998).
The ionization state of the absorber remained largely unknown.
However, there is now growing evidence that these absobers
are not cold but warm. 
As shown by Schartel et al. (1997) the spectrum and spectral changes of 
the high-redshift quasar PKS\,2351-154 
($z$=2.67)
are well
explained by the presence of an ionized absorber
of column density $\log N_{\rm w} = 22.4$ which changes its ionization state
in response to intrinsic luminosity changes of the quasar.
PKS\,2351-154 is one of the very few high-$z$ quasars which show a {\em variable} UV
  absorption system as well.
For several years, this quasar 
 held the record of being the most distant X-ray
warm-absorber candidate known,
recently exceeded by GB\,1428+42 and PMN\,J0525-33 (Fabian et al. 2001a,b).
In contrast, Yuan et al. (2000) argued that the X-ray absorber
of the high-redshift quasar RXJ1028.6-0844 is very likely cold.
An interesting  puzzle is, why the UV
spectrum of this object does not show any signs of the X-ray cold absorber.
 
\begin{figure}[h]
\hspace*{0.3cm}
\psfig{file=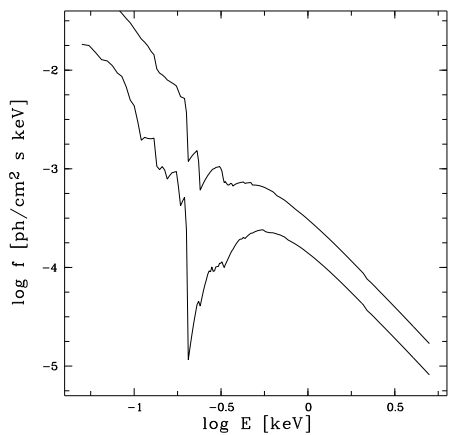,width=6.0cm,clip=}
\hfill
\begin{minipage} []{0.57\hsize}\vspace*{-4.7cm}
\hfill
\caption{ Best-fit warm absorber which was used to model the X-ray
spectrum of the $z=2.69$ quasar PKS\,2351-154
in high-state (upper curve) and low-state (lower curve; Schartel et al. 1997). 
For even higher redshifts, edges of
the ionized absorber are shifted ever closer to the
low-energy sensitivity cut-off of current X-ray instruments.
In case of insufficient S/N, the edges and lines of the warm absorber
may easily be confused with cold hydrogen absorption, and thus
mimic the presence of a {\em cold} absorber of high column density
(Komossa, e.g., 1999).  
}
\end{minipage} 
\label{wa}
 \end{figure}

\subsection{Open questions which will be addressed with {\sl XEUS}}

What is the origin and nature of the high-z excess absorbers ?
Is this material warm or cold ?  
Why has it been more abundant in the past ? How does it evolve ?
  Why is excess absorption
mainly seen in high-redshift radio-{\em loud} quasars
whereas a number of (non-BAL) high-z radio-{\em quiet}
quasars appear to be absorption-free ?  Answers to these questions
are crucial for understanding the formation and evolution of AGN. 

Apart from measuring ionic column densities, a very interesting
prospect is to determine element abundances in dust and gas 
at high redshift.  
This topic will be discussed in more
detail in Section 6. 

\section{\large Absorption lines from the intergalactic medium} 

An exciting new aspect of having access to high-resolution spectroscopy
in X-rays
is the search for absorption lines from the intergalactic
medium. 
First results were recently reported by Mathur et al. (2002).
(Weak) OVII and OVIII absorption lines in the direction
of the bright quasar H1821+643 were interpreted as arising
from a moderate density, shock-heated intergalactic medium
predicted by cosmological scenarios. 
(See Fang et al. 2002 for detection of OVIII Lyman-alpha absorption
from intra-group gas along the sightline towards PKS\,2155-304). 

With {\sl XEUS}, we will be able to measure more lines of sight, and greatly
improve the statistics, to measure reliably high-ionization lines from filaments
of low column density.

\section{\large Prospects for studying absorbers at high redshift with {\sl{XEUS}}:
              constraints on early star-formation history and cosmological parameters
from iron abundance measurements}

The spectral richness and complexity of AGN, observed
with {\sl Chandra} and {\sl XMM},  will provide a
wealth of information on the physical processes in the
central region of the few X-ray brightest, most nearby Seyfert galaxies.
The long-term goal is to obtain this information for a much
larger number of objects, particularly at higher redshift.

We concentrate here on the aspect of deriving metal abundances 
at high redshift. 
Below, we summarize open questions and how they can be addressed
with the greatly improved sensitivity and resolution (Arnaud et al. 1999)
of {\sl XEUS}.

\subsection{Abundances at high redshift: constraints on early star-formation history of the
universe, and on cosmological parameters}

Two types of quasars show excess absorption at high redshift:
radio-loud quasars on the one hand, BAL quasars on the other
hand.  
((Interestingly, though, very few BALs are radio-loud;
but whether these facts tell us something about the
similar/different origin of the excess absorbers in both types 
of objects, is unclear.)) 

Here, we will 
concentrate on the iron {\em edge} features
which offer some valuable advantages  over Fe emission lines
in usage as abundance indicators
(for interesting aspects of studying relativistically
 broadened iron {\em lines} at high redshift 
we refer to the {\sl XEUS science case}; Arnaud et al. 1999). 

The iron absorption features and K-edges provide a unique
probe of matter at high redshift because, firstly, they are easy 
to measure even, or, {\em particularly}, at high redshift $z$,
and secondly, Fe(/O) abundance measurements in the early
universe are important for key issues of cosmology, as elaborated on
below.   
The relevance of UV-FeII emission in deriving Fe/Mg abundances 
at high $z$ was discussed by 
Hamann \& Ferland (1999) and Yoshii et al. (1998).
In particular, Yoshii et al. (1998) used the emission-line ratio
FeII/MgII to estimate the iron-to-magnesium abundance in the 
distant quasar QSO\,B1422+231.  
Here, we would like to emphasize the role  
of the X-ray FeK$\alpha$ edges as one of the `cleanest' diagnostics
of iron -- of column density and ionization state in a first step,
and of Fe/O abundance in a second step:

\begin{itemize}

\item Firstly, the iron absorption is easy to measure:
At high redshift, the iron-K edges are redshifted to
a region, where the detector sensitivity is high. 
As a bonus, the continuum shape is very well constrained
since redshift shifts the high-energy part of
the quasar spectrum into the observable band. \\ 

\item Absorption edges are, in principle, more reliable
than emission lines because they do not depend
on parameters like gas density and temperature, 
and provide a direct measurement of the ionic column densities.

In addition, the iron edge is better suited than
the iron K$\alpha$ emission line which appears to be less
common at high redshift.  \\ 

In the UV-band, no strong Fe-lines are present,
apart from the FeII emission complexes which 
are still subject to high uncertainties,
(i) in model predictions (e.g., the role of photo-excitation by
Lyman $\alpha$ photons),  
and (ii) in measurement interpretation (some Fe may be depleted
into dust, some Fe may be of higher ionization state than  
FeII, with reduced contribution to FeII emission  in both cases);
see Hamann \& Ferland (1999) and references therein.  \\

\item The iron edges do not coincide with any other strong
absorption-line transitions at the same energies
(as opposed to some low-energy features, where K-shell
and L- and M-shell features of different species 
partly overlap). \\

\item Absorbers of {\em high column density} have been
observed in high-redshift quasars in X-rays,
particularly in radio-loud quasars and BAL quasars,
so they are known to be present. A high column density
is indeed required for the optical depths in the edges
to become measurable. E.g., a column density
in neon-like iron of $N_{\rm FeXVII} = 10^{19}$ cm$^{-2}$
corresponds to an optical depth of $\tau = 0.27$
in the absorption edge 
 (with cross section from Verner \& Yakovlev 1995).   \\ 

\item The element iron plays a special role in 
chemical enrichment scenarios, because its production
is delayed relative to other elements (e.g., Fig 1 of Hamann
\& Ferland 1993), since it is believed to be mostly produced
in supernovae of type Ia (e.g., 
Nomoto et al. 1984, Sect. 2.1 of Greggio \& Renzini, 
1984, Sect.4.6 of Hamann \& Ferland 1993, and references therein). 
Its role as ``cosmic clock'' was therefore realized early (e.g., Tinsley 1979).
Hamann \& Ferland (1993) also pointed out its role in determining
cosmological parameters: A certain age of the universe is required
to produce iron in sufficient amounts.
The detection of high Fe abundances at high redshift would 
therefore point to a larger age of the universe at the same
redshift, thus to a different set of cosmological parameters
than 
an Einstein-deSitter universe with deceleration parameter
$q_o=0.5$ (Fig. 4).

\end{itemize}

It is these last two science issues, which 
are of special interest.
The usage of iron as a ``cosmic clock'' depends on our understanding
of supernovae of type Ia, which are thought to play the
dominant role in the enrichment of iron relative to alpha 
elements.  
Given the long lifetime of SN\,Ia precursors,
it takes about 0.3-1\,Gyr until SN Ia
start to form in significant numbers 
(e.g., Fig. 9 of Hamann \& Ferland 1993; Fig. 4 of Matteucci 1994).
The 
iron production is delayed correspondingly ($t >$ 1Gyr). 
Therefore, even with a high rate of SN\,Ia, detection of large amounts
of iron in the very early universe would require another
mechanism to be at work.
It would most likely imply a larger age of the universe at a given
redshift, to provide more time for the formation of iron
(Hamann \& Ferland 1993, Yoshii et al. 1998).  

\begin{figure}[hb]
\hspace*{0.3cm}
\psfig{file=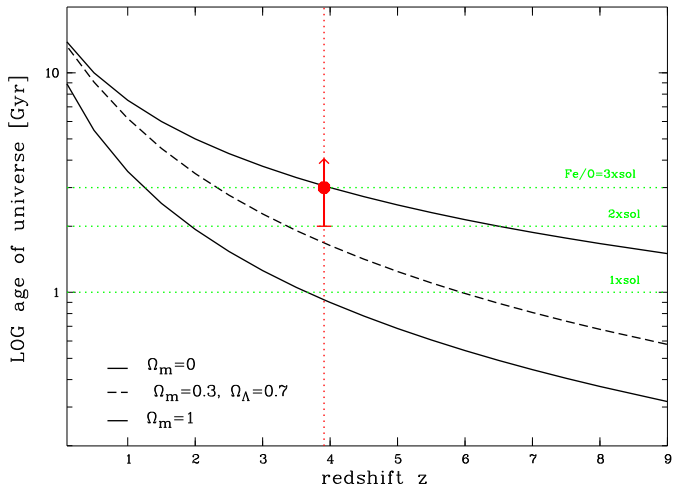,width=9.0cm,clip=}
\hfill
\begin{minipage}[]{0.41\hsize}\vspace*{-7.3cm}    
\hfill
\caption{
{\footnotesize{Age of the universe in units of 10$^{9}$ 
years versus redshift for different cosmologies,
using a Hubble constant $H_{\rm 0}=65$ km/s/Mpc.
Upper solid line: $\Omega_{\rm m}$=0, 
lower solid line: $\Omega_{\rm m}$=1 ($\Omega_{\rm \Lambda}$=0 in both cases),
dashed line: $\Omega_{\rm m}$=0.3 and $\Omega_{\rm \Lambda}$=0.7.
The dotted horizontal lines correspond to the timescale to
produce a Fe/O abundance ratio of solar, 2 and 3$\times$ solar
(Hamann \& Ferland 1993; their model `M4a'),
as marked in the figure. The vertical dotted line corresponds
to a redshift $z=3.91$ measured for APM\,08279+5255.
The filled circle with the error bar gives the timescale
necessary to produce the observed amount of iron in APM\,08279+5255
(based on model M4a of Hamann \& Ferland). {\sl XEUS} will be crucial
in determining Fe abundances at high redshift. It will allow to significantly
narrow down the error bars for APM\,08279+5255, and to study
more high-redshift objects in a similar way. 
}}
}
\end{minipage}
 \end{figure}
\begin{table}[h]
\begin{tabular}{llllll}
  \noalign{\smallskip}
  \hline
  \noalign{\smallskip}
cosmological model & age of universe & \multicolumn{3}{l}{age of universe at $z$=} & \\
  \noalign{\smallskip}
                   & $t_{\rm look-back,z\rightarrow \inf}$ & 9.0 & 6.0 & 3.9 & 1.0 \\
  \noalign{\smallskip}
  \hline
  \noalign{\smallskip}
$\Omega_{\rm m}$=0, $\Omega_{\rm \Lambda}$=0 & 15 Gyr   & 1.5 Gyr & 2.1 Gyr & 3.0 Gyr & 7.5 Gyr \\
$\Omega_{\rm m}$=0.3, $\Omega_{\rm \Lambda}$=0.7 & 14.5 & 0.6     & 1.0    & 1.7    & 6.1   \\
$\Omega_{\rm m}$=1, $\Omega_{\rm \Lambda}$=0 & 10       & 0.3     & 0.5    & 0.9    & 3.5   \\
  \noalign{\smallskip}
  \hline

\end{tabular}
\hfill
\begin{minipage}[]{0.3\hsize}\vspace*{1.3cm}   
\hfill
\caption{Age of the universe (in Giga years) at different redshifts for several cosmologies.} 
\end{minipage}

\end{table}

To demonstrate these issues further, we use the recent
X-ray results of Hasinger et al. (2002) on the BAL quasar 
APM 08279+5255 at redshift $z$=3.91,
which shows an Fe/O ratio of about 3.
Such high ratios are not produced by SN\,II/Ib and imply
that SN\,Ia are involved. However, assuming 
an Einstein-deSitter world model with  
$\Omega_{\rm m}$=1.0 and $\Omega_{\rm \Lambda}$=0,  
at the redshift
of the quasar the universe was too young ($t \approx$ 0.9\,Gyr; Tab. 1)
to produce the observed overabundance of iron.
According to models of Hamann \& Ferland (1993; their model M4a)
a timescale of $\sim$3 Gyr is required to produce an
abundance ratio of Fe/O=3.{\footnote{The timescale to
reach an abundance ratio of Fe/O=solar is
{\em at least} 1 Gyr in all models of Hamann \& Ferland (HF93),
and is basically given by the lifetime of SNIa precursors. 
Different models of  HF93 then predict a different evolution
of F/O, and we caution that in model predictions there is some scatter
in the time at which Fe/O reaches 3$\times$solar. 
Model `M4' is the quasar model favored by HF93. In several other models
they studied, Fe/O never reaches 3$\times$solar, whereas in their
extreme model `M6' (cf. their Fig. 1) Fe/O reaches 3$\times$solar
already after 2 Gyr.}} 
 
The {\sl XMM-Newton} X-ray observations of APM 08279+5255
therefore favor cosmological models which
predict a slightly larger age of the universe {\em at the redshift
of the quasar}, like recent models involving a low
value of $\Omega_{\rm matter}$ and a cosmological
constant (e.g., Fig. 7 of Perlmutter et al. 1999). 
The idea is illustrated in Fig. 4,
which plots the age of the universe in dependence of redshift $z$ 
(e.g., Carrol et al. 1992)
for several cosmological models. 

The excellent spectral resolution and sensitivity of {\sl XEUS}
will not only allow us to scrutinize the presence of the iron feature 
in APM 08279+5255 and its interpretation, it will also provide
similar information for many more objects. We will thus obtain 
valuable information about nucleosynthesis in the early universe,
and we will be able to follow another path to measure cosmological
parameters.

\subsection{Soft X-ray sensitivity of future X-ray missions}

An important design feature of future X-ray missions, particularly
those aiming at high-redshift studies, is the {\em soft} X-ray
sensitivity. In order to determine metal abundance ratios of Fe/O
and Fe/Ne, it will be essential
to detect oxygen and neon edges out to as high redshift as
possible{\footnote{For that purpose, the edges have to be disentangled
from other potential absorption lines at similar energies,
and any potential black body contribution at the softest X-ray energies
has to be measured carefully, since it can influence the 
oxygen-iron ionization structure, thus affecting the measured
ratio of Fe/O and Fe/Ne (Hasinger, Komossa et al., in prep.).
The possibility of partial re-filling of absorption features
due to scattering also has to be kept in mind carefully.}}.  
The high sensitivity of {\sl XEUS} in its final configuration
will be crucial to study iron absorption features at high redshifts, 
since the iron abundance is expected to decline
significantly beyond a redshift $z \approx$ 4, as discussed above.

\section{\large Summary}
Science issues that we will be able to address with {\sl XEUS}
for the first time, particularly at high redshift, include:
measurement of metal abundances
of X-ray (cold) absorbers by
detection of metal absorption edges,
determination of the composition of dust mixed with cold and ionized
gas (K-edges of metals in cold dust and cold gas will be resolvable from each other
 for the first time),
 measurement of the velocity state of the gas,
and usage of these results to investigate the {\em evolution} of gas and dust
in AGN from high to low redshift: the evolution of abundances, dust content,
ionization state,
amount and velocity of gas, and its role in feeding the black hole.

A particularly exciting aspect is to use the iron 
absorption features, especially the K-edges, for
abundance determinations, which will have fundamental
implications for our understanding of 
the early star formation history of the universe,
and which will provide another means to
measure cosmological parameters.  
In order to determine Fe/O and Fe/Ne ratios out to large redshifts, 
{\em soft} X-ray energy sensitivity (to reliably measure O, Ne)
of future X-ray missions is essential.

\begin{acknowledgements}
StK thanks Hartmut Schulz for very useful discussions
on cosmological models.  
\end{acknowledgements}

\end{document}